# Influence of String Stiffness on Piano Tone

Lai-Mei Nie

*Department of Physics,
Tsinghua University,
Beijing 100084, China*

**Abstract**：Piano tones vary according to how pianist touches the keys. Many possible factors contribute to the relations between piano touch and tone. Focusing on the stiffness of string, we establish a model for vibration of a real piano string and derive a semi-analytical solution to the vibration equation.

From soft and dim to bright and sharp, modern piano has a variety of tones, which greatly contributes to its popularity over three hundred years. However, physical description of piano acoustics came relatively late due to complexity of mechanical structure[1]. Based on numerical solutions of vibration equation of piano string and acoustical measurements, some solid conclusions have been obtained about the properties of piano mechanics and acoustics[2-5]. In this article, instead, we provide a semi-analytical approach to explore the relationship between piano touch and tone. Especially, we focus on the influence of string stiffness.

The article consists of two sections. In Sec.1, a physical model of piano string vibration is established and vibration equation is obtained. Sec.2 gives the semi-analytical solution to the equation.

## 1. The model and vibration equation

Our starting point is to consider tension force and shear force on the string, as shown in Fig. 1.1, where $z(x,t)$ is transverse displacement of the string. We have

$$\rho \cdot S \cdot \Delta x \cdot \frac{\partial^2 z}{\partial t^2} = [F + \frac{\partial F}{\partial x} \cdot \Delta x] - F + T_1 \sin\alpha_1 - T_2 \sin\alpha_2, \tag{1.1}$$

where $\rho$ is the mass density of string, $S$ is the area of cross section of the string. Due to small amplitude of vibration, $T_1 \approx T_2 \equiv T$, and Eq. (1.1) becomes

$$\rho \cdot S \cdot \frac{\partial^2 z}{\partial t^2} = \frac{\partial F}{\partial x} + T \cdot \frac{\partial^2 z}{\partial x^2} \tag{1.2}$$

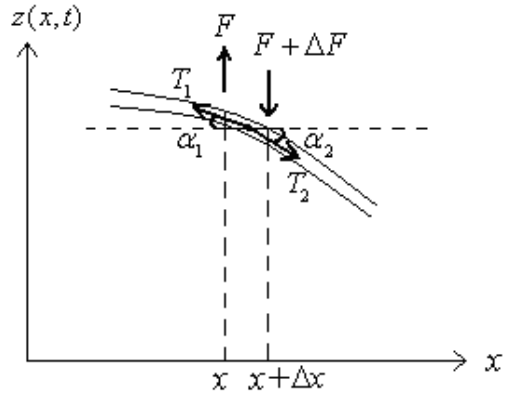

Fig. 1.1　String displacement under tension force $T$ and shear force $F$.

The relation between shear force $F$ and bending moment $M$ is

$$F = -\frac{\partial M(x)}{\partial x}. \tag{1.3}$$

We need the relation between $M$ and $z(x,t)$. Effect of shear force is presented as bending moment on the string, while effect of tension force is to pull the string. Thus, we separate these two effects into two steps: first increase the tension in the string and then bend it into final configuration, as shown in Fig. 1.2.

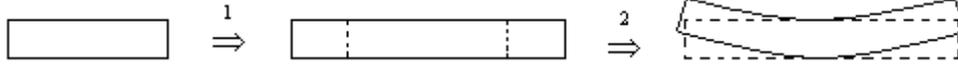

Fig. 1.2.  String deformation under tension and shear force.

After Step 1 and Step 2, $\widehat{o_1 o_2}$ and $\widehat{n_1 n_2}$ stretch, as shown in Fig. 1.3, where the neutral layer $\widehat{o_1 o_2}$ does not change its length under pure bending of Step 2.

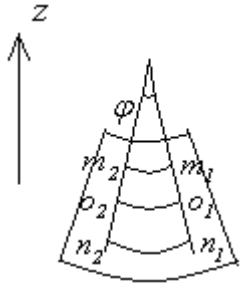

Fig. 1.3  Deformation of a string segment. Note that $\widehat{o_1 o_2}$ does not change its length under pure bending.

For $\widehat{m_1 m_2}$, the uniaxial strain produced by Step 1 is

$$\varepsilon_{m_1} = \frac{T}{ES}, \tag{1.4}$$

where $E$ is Young's modulus. The uniaxial strain of $\widehat{m_1 m_2}$ produced by Step 2 is

$$\varepsilon_{m_2} = \frac{\widehat{m_1 m_2} - \widehat{o_1 o_2}}{\widehat{o_1 o_2}} = \frac{(R-z)\varphi - R\varphi}{R\varphi} = -\frac{z}{R}, \qquad (1.5)$$

where $R$ is radius of curvature and $\widehat{o_1 o_2}$ locates at $z = 0$. Then the total uniaxial strain of $\widehat{m_1 m_2}$ is

$$\varepsilon_m = \varepsilon_{m_1} + \varepsilon_{m_2} = \frac{T}{ES} - \frac{z}{R}. \qquad (1.6)$$

Similarly, we have the uniaxial strains of $\widehat{o_1 o_2}$ and $\widehat{n_1 n_2}$ as follows:

$$\varepsilon_o = \varepsilon_{o_1} + \varepsilon_{o_2} = \frac{T}{ES} + 0 = \frac{T}{ES} - \frac{z}{R} \quad (z=0), \qquad (1.7)$$

$$\varepsilon_n = \frac{T}{ES} - \frac{z}{R}. \qquad (1.8)$$

Thus we have

$$\varepsilon = \frac{T}{ES} - \frac{z}{R} \qquad (1.9)$$

for $\widehat{m_1 m_2}$, $\widehat{o_1 o_2}$ and $\widehat{n_1 n_2}$. Therefore,

$$\begin{aligned} M(x) &= -\int z \cdot \varepsilon \cdot dS \\ &= -\frac{T}{S}\int z dS + \frac{E}{R}\int z^2 dS \\ &= \frac{EJ}{R}, \end{aligned} \qquad (1.10)$$

where $J = \int z^2 dS$ is the second moment of area. Substituting

$$\frac{1}{R} = \frac{\frac{\partial^2 z}{\partial x^2}}{[1+(\frac{\partial z}{\partial x})^2]^{\frac{3}{2}}} \qquad (1.11)$$

into Eq.(1.10) and considering small vibration, i.e., $\left|\frac{\partial z}{\partial x}\right| \ll 1$, we have

$$M(x) = EJ \cdot \frac{\partial^2 z}{\partial x^2}. \qquad (1.12)$$

From Eqs. (1.2), (1.3) and (1.12), we arrive at the vibration equation

$$\frac{\partial^2 z}{\partial t^2} + \frac{EJ}{\rho S}\frac{\partial^4 z}{\partial x^4} - \frac{T}{\rho S}\frac{\partial^2 z}{\partial x^2} = 0. \tag{1.13}$$

## 2. Semi-analytical solution of vibration equation.

We use the approach of separation of variables to solve Eq. (1.13). Set

$$z(x,t) = X(x)f(t), \tag{2.1}$$

then we have two ordinary differential equations:

$$\frac{d^2 f}{dt^2} + \lambda f(t) = 0, \tag{2.2}$$

$$\frac{EJ}{\rho S} \cdot \frac{d^4 X}{dx^4} - \frac{T}{\rho S}\frac{d^2 X}{dx^2} - \lambda X(x) = 0, \tag{2.3}$$

where $\lambda$ is a constant to be determined from boundary conditions. From Eq. (2.2) we have

$$f(t) = A\sin(\sqrt{\lambda}t) + B\cos(\sqrt{\lambda}t), \tag{2.4}$$

where we have set $\lambda > 0$ which represents for angular frequency.

We now solve Eq. (2.3). Its characteristic equation is

$$\frac{EJ}{\rho S} \cdot x^4 - \frac{T}{\rho S} \cdot x^2 - \lambda = 0. \tag{2.5}$$

Setting $y = x^2$, we get

$$\frac{EJ}{\rho S} \cdot y^2 - \frac{T}{\rho S} y - \lambda = 0. \tag{2.6}$$

Let $y_1, y_2$ be the two real roots of the equation above, and $y_1 > 0 > y_2$, $|y_1| > |y_2|$. Then the solution of Eq. (2.6) is

$$\begin{cases} x_1 = \sqrt{y_1} \\ x_2 = -\sqrt{y_1} \\ x_3 = i\sqrt{-y_2} \\ x_4 = -i\sqrt{-y_2}, \end{cases} \tag{2.7}$$

where $i$ is the imaginary unit. For convenience, set

$$\begin{cases} \tilde{x}_1 = x_1 = \sqrt{y_1} \\ \tilde{x}_2 = x_2 = -\sqrt{y_1} \\ \tilde{x}_3 = \sqrt{-y_2} \\ \tilde{x}_4 = -\sqrt{-y_2} \end{cases} \tag{2.8}$$

then the solution of Eq. (2.3) is

$$X(x) = C_1 e^{\tilde{x}_1 x} + C_2 e^{\tilde{x}_2 x} + (C_3^{(1)} \cos \tilde{x}_3 x + C_3^{(2)} \sin \tilde{x}_3 x) + (C_4^{(1)} \cos \tilde{x}_4 x + C_4^{(2)} \sin \tilde{x}_4 x)$$

$$= C_1 e^{\tilde{x}_1 x} + C_2 e^{\tilde{x}_2 x} + C_3 \cos \tilde{x}_3 x + C_4 \sin \tilde{x}_3 x \tag{2.9}$$

The boundary conditions

$$X(0) = 0, \ X(L) = 0, \ X'(0) = 0, \ X'(L) = 0 \tag{2.10}$$

lead to the secular equation

$$\mathrm{Det} \begin{pmatrix} 1 & 1 & 1 & 0 \\ e^{\sqrt{y_1} L} & e^{-\sqrt{y_1} L} & \cos\sqrt{-y_2} L & \sin\sqrt{-y_2} L \\ \sqrt{y_1} & -\sqrt{y_1} & 0 & \sqrt{-y_2} \\ \sqrt{y_1} e^{\sqrt{y_1} L} & -\sqrt{y_1} e^{-\sqrt{y_1} L} & -\sqrt{-y_2} \sin\sqrt{-y_2} L & \sqrt{-y_2} \cos\sqrt{-y_2} L \end{pmatrix} = 0, \tag{2.11}$$

where $L$ is the length of string.

Now we need the values of parameters such as $L$, $\dfrac{EJ}{\rho S}$ and $\dfrac{T}{\rho S}$ to solve $\lambda$ from Eq. (2.11).

Instead of using precise values of a certain piano, we consider a group of approximate values of parameters to obtain generality[6]:

$$\begin{cases} L = 0.4 m \\ E = 200 \times 10^9 Pa \\ \rho S = 0.005 kg/m \\ T = 900 N \\ J = 0.25\pi R^4 = 0.25\pi (0.45 \times 10^{-3})^4 (m)^4. \end{cases} \tag{2.12}$$

Set $E_1 \equiv \dfrac{EJ}{\rho S} = 1.288 \dfrac{N \cdot m^3}{kg}$, $T_1 \equiv \dfrac{T}{\rho S} = 1.800 \times 10^5 \dfrac{N \cdot m}{kg}$, and Eq. (2.6) becomes

$$E_1 y^2 - T_1 y - \lambda = 0, \tag{2.13}$$

together with its solutions

$$y_1 = \frac{T_1 + \sqrt{T_1^2 + 4E_1\lambda}}{2E_1}, \quad y_2 = \frac{T_1 - \sqrt{T_1^2 + 4E_1\lambda}}{2E_1}. \quad (2.14)$$

Setting

$$f(\lambda) = \text{Det} \begin{pmatrix} 1 & 1 & 1 & 0 \\ e^{\sqrt{y_1}L} & e^{-\sqrt{y_1}L} & \cos\sqrt{-y_2}L & \sin\sqrt{-y_2}L \\ \sqrt{y_1} & -\sqrt{y_1} & 0 & \sqrt{-y_2} \\ \sqrt{y_1}e^{\sqrt{y_1}L} & -\sqrt{y_1}e^{-\sqrt{y_1}L} & -\sqrt{-y_2}\sin\sqrt{-y_2}L & \sqrt{-y_2}\cos\sqrt{-y_2}L \end{pmatrix}, \quad (2.15)$$

we use Mathematica to find roots of $f(\lambda) = 0$. The results are (with the unit $\frac{N}{m \cdot kg}$):

$$\begin{cases} \lambda_1 = 1.14 \times 10^7 \\ \lambda_2 = 4.57 \times 10^7 \\ \lambda_3 = 1.03 \times 10^8 \\ \lambda_4 = 1.84 \times 10^8 \\ \lambda_5 = 2.88 \times 10^8 \\ \lambda_6 = 4.17 \times 10^8 \\ \lambda_7 = 5.71 \times 10^8, \text{ etc.} \end{cases} \quad (2.16)$$

where $\lambda_1$ belongs to the fundamental wave, $\lambda_2$ belongs to the first harmonic and so on. Taking $\lambda_1$ as an example, we substitute it into the linear homogeneous equations

$$\begin{pmatrix} 1 & 1 & 1 & 0 \\ e^{\sqrt{y_1}L} & e^{-\sqrt{y_1}L} & \cos\sqrt{-y_2}L & \sin\sqrt{-y_2}L \\ \sqrt{y_1} & -\sqrt{y_1} & 0 & \sqrt{-y_2} \\ \sqrt{y_1}e^{\sqrt{y_1}L} & -\sqrt{y_1}e^{-\sqrt{y_1}L} & -\sqrt{-y_2}\sin\sqrt{-y_2}L & \sqrt{-y_2}\cos\sqrt{-y_2}L \end{pmatrix} \begin{pmatrix} C_1 \\ C_2 \\ C_3 \\ C_4 \end{pmatrix} = 0 \quad (2.17)$$

and obtain

$$\begin{pmatrix} 1 & 1 & 1 & 0 \\ e^{149.567} & 1/e^{149.567} & \cos(3.1842) & \sin(3.1842) \\ 373.918 & -373.918 & 0 & 7.9604 \\ 373.918e^{149.567} & -373.918e^{-149.567} & -7.9604 \times \sin 3.1842 & 7.9604 \times \cos 3.1842 \end{pmatrix} \begin{pmatrix} C_1 \\ C_2 \\ C_3 \\ C_4 \end{pmatrix} = 0$$

$$(2.18)$$

This is a badly conditioned matrix and cannot be solved unless we use some approximation. We notice that $C_1$ should to be extremely small to keep its product finite with large coefficients. Therefore, Eq.(2.18) is equivalent to the following equations:

$$C_2 + C_3 = 0, \tag{2.19}$$

$$C_1 e^{\tilde{x}_1 L} + C_2 e^{\tilde{x}_2 L} + C_3 \cos \tilde{x}_3 L + C_4 \sin \tilde{x}_3 L = 0, \tag{2.20}$$

$$\tilde{x}_2 C_2 + \tilde{x}_3 C_4 = 0, \tag{2.21}$$

$$\tilde{x}_1 C_1 e^{\tilde{x}_1 L} + \tilde{x}_2 C_2 e^{\tilde{x}_2 L} - \tilde{x}_3 C_3 \sin \tilde{x}_3 L + \tilde{x}_3 C_4 \cos \tilde{x}_3 L = 0, \tag{2.22}$$

where we have taken $C_1 = 0$ in Eqs. (2.19) and (2.21), and preserved it in Eqs. (2.20) and (2.22). We have adopted Eq. (2.8) for convenience. Taking $C_2 = 1$, we obtain

$$C_1 = \frac{\cos \tilde{x}_3 L - e^{-\tilde{x}_1 L} - \dfrac{\tilde{x}_1}{\tilde{x}_3} \sin \tilde{x}_3 L}{e^{\tilde{x}_1 L}}, \tag{2.23}$$

$$C_3 = -1, \tag{2.24}$$

$$C_4 = \frac{\tilde{x}_1}{\tilde{x}_3}. \tag{2.25}$$

So far, we have obtained the coefficients $C_1 \sim C_4$ for the solution of vibration equation:

$$z(x,t) = X(x) f(t) = \sum_n X_n(x) \left[ A_n \sin \sqrt{\lambda_n} t + B_n \cos \sqrt{\lambda_n} t \right], \tag{2.26}$$

where $X_n(x) = C_1^{(n)} e^{\tilde{x}_1^{(n)} x} + C_2^{(n)} e^{\tilde{x}_2^{(n)} x} + C_3^{(n)} \cos \tilde{x}_3^{(n)} x + C_4^{(n)} \sin \tilde{x}_3^{(n)} x$ belonging to $\lambda_n$.

From the initial condition $z(x, 0) = 0$, we have $B_n = 0$, and thus

$$z(x,t) = \sum_n A_n \cdot X_n(x) \cdot \sin \sqrt{\lambda_n} t$$

$$= \sum_n A_n \sin \sqrt{\lambda_n} t \cdot \left( C_1^{(n)} e^{\tilde{x}_1^{(n)} x} + C_2^{(n)} e^{\tilde{x}_2^{(n)} x} + C_3^{(n)} \cos \tilde{x}_3^{(n)} x + C_4^{(n)} \sin \tilde{x}_3^{(n)} x \right) \tag{2.27}$$

Now we will fix $A_n$ to see how $\dfrac{A_n}{A_1}$ varies when hitting force increases. From another initial condition

$\left. \dfrac{\partial z(x,t)}{\partial t} \right|_{t=0}$, we get

$$A_n = \dfrac{\dfrac{1}{\sqrt{\lambda_n}} \int_0^L \left. \dfrac{\partial z(x,t)}{\partial t} \right|_{t=0} \cdot X_n(x) \mathrm{d}x}{\int_0^L [X_n(x)]^2 \mathrm{d}x}, \qquad (2.28)$$

with the proof of orthogonal eigenfunctions in Appendix. We take $n=7$ as an example:

$$\dfrac{A_7}{A_1} \propto \dfrac{\int_0^L \left. \dfrac{\partial z(x,t)}{\partial t} \right|_{t=0} \cdot X_7(x) \mathrm{d}x}{\int_0^L \left. \dfrac{\partial z(x,t)}{\partial t} \right|_{t=0} \cdot X_1(x) \mathrm{d}x}$$

$$= \dfrac{\int_0^L \left. \dfrac{\partial z(x,t)}{\partial t} \right|_{t=0} \left[ C_1^{(7)} e^{\tilde{x}_1^{(7)} x} + C_2^{(7)} e^{\tilde{x}_2^{(7)} x} + C_3^{(7)} \cos \tilde{x}_3^{(7)} x + C_4^{(7)} \sin \tilde{x}_3^{(7)} x \right] \mathrm{d}x}{\int_0^L \left. \dfrac{\partial z(x,t)}{\partial t} \right|_{t=0} \left[ C_1^{(1)} e^{\tilde{x}_1^{(1)} x} + C_2^{(n)} e^{\tilde{x}_2^{(1)} x} + C_3^{(1)} \cos \tilde{x}_3^{(1)} x + C_4^{(1)} \sin \tilde{x}_3^{(1)} x \right] \mathrm{d}x} \qquad (2.29)$$

We take the initial condition $\left. \dfrac{\partial z(x,t)}{\partial t} \right|_{t=0}$ to be the simple form:

$$\left. \dfrac{\partial z(x,t)}{\partial t} \right|_{t=0} \equiv v(x) = \begin{cases} 0, & 0 \le x \le c-d \\ b_0, & c-d \le x \le c+d, \\ 0, & c+d \le x \le L \end{cases} \qquad (2.30)$$

where $c$ is the point where hammer hits the string, $[c-d, c+d]$ is their contact range, as shown in Fig. 2.1.

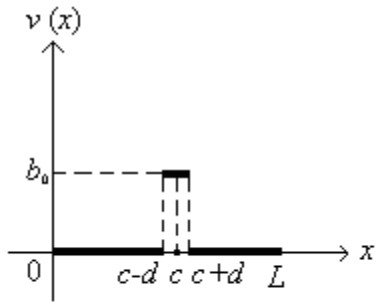

Fig. 2.1. Initial condition of string when hit by a hammer

Using $\lambda_1 = 1.14 \times 10^7 \, \dfrac{N}{m \cdot kg}$, we obtain (with the unit $m$)

$$\begin{cases} C_1^{(1)} = 1.106 \times 10^{-65} \\ C_2^{(1)} = 1 \\ C_3^{(1)} = -1 \\ C_4^{(1)} = 46.972. \end{cases} \tag{2.31}$$

Similarly, we can get for $\lambda_7 = 5.71 \times 10^7 \dfrac{N}{m \cdot kg}$:

$$\begin{cases} C_1^{(7)} = 2.194 \times 10^{-66} \\ C_2^{(7)} = 1 \\ C_3^{(7)} = -1 \\ C_4^{(7)} = 6.784. \end{cases} \tag{2.32}$$

Thus,

$$\frac{A_7}{A_1} \propto \frac{\int_0^L \left.\frac{\partial z(x,t)}{\partial t}\right|_{t=0} \cdot X_7(x)\mathrm{d}x}{\int_0^L \left.\frac{\partial z(x,t)}{\partial t}\right|_{t=0} \cdot X_1(x)\mathrm{d}x} \propto \frac{\eta_7}{\eta_1} \equiv G(c,d), \tag{2.33}$$

where

$$\eta_i \equiv \frac{C_1^{(i)}}{\tilde{x}_1^{(i)}}\left[\exp[\tilde{x}_1^{(i)}(c+d)] - \exp[\tilde{x}_1^{(i)}(c-d)]\right] + \frac{C_2^{(i)}}{\tilde{x}_1^{(i)}}\left[\exp[\tilde{x}_1^{(i)}(c+d)] - \exp[\tilde{x}_1^{(i)}(c-d)]\right]$$

$$+ \frac{C_4^{(i)}}{\tilde{x}_3^{(i)}}\{\cos[\tilde{x}_3^{(i)}(c-d)] - \cos[\tilde{x}_3^{(i)}(c+d)]\} + \frac{C_3^{(i)}}{\tilde{x}_3^{(i)}}\{\sin[\tilde{x}_3^{(i)}(c+d)] - \sin[\tilde{x}_3^{(i)}(c-d)]\}, \quad i = 1, 7.$$

(2.34)

In contemporary pianos, $0.08 \leq \dfrac{c}{L} \leq 0.12$ [5]. When $L = 0.4m$, we have $0.032m \leq c \leq 0.048m$. Especially, $c \sim 0.048m$ in the middle range[5]. We plot $G$ ($c=0.048,d$) in Fig. 2.2 (see next page) as a function of $d$, which increases with the increment of hitting force on the key.

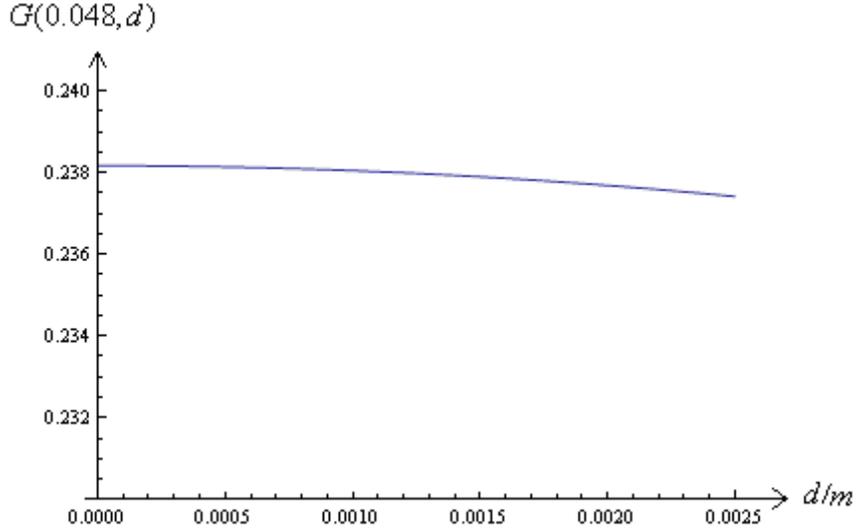

Fig. 2.2.   $G(c=0.048,d)$ as a function of the contact range $d$.

From Fig. 2.2, we see that when hitting the key more strongly, we would get more dim tones due to the stiffness of piano string. This conclusion seems to contradict with our daily experience that tone of piano becomes bright as the hitting force exerted on the keys increases. This paradox may come from other reasons. Here we list three of them.

(1) The time duration that hammer and string contact may be a very crucial factor. The softer we hit a key, the longer the hammer and string interact with each other, which may suppress higher frequencies of vibration and lead to dimmer tone.

(2) We have taken the initial conditions to have the simplest form, as shown in Eq. (2.30). However, the interaction between hammer and string can be much more complex than that. For example, the velocity varies along the string and is not a constant within the contact range, and the force hammer exerted on string is not instantaneous, which should be taken into account as a time-dependent source in the vibration equation. It is supposed to apply numerical methods under this condition[2].

(3) What we have focused on is the hammer-string system in our research. Our analysis would be more close to reality if we take into account the vibration of soundboard. Due to the complexity of soundboard, such as shape and material, this analysis calls for precise experiments and computer simulations.

## Appendix

Here we provide the proof of orthogonality of $X_n(x)$ which belong to different eigenvalues $\lambda_n$.

When $\lambda_2 \neq \lambda_1$,

$$\frac{EJ}{\rho S} X_1'''' - \frac{T}{\rho S} X_1'' - \lambda_1 X_1 = 0, \qquad (Ap.1)$$

$$\frac{EJ}{\rho S}X_2'''' - \frac{T}{\rho S}X_2'' - \lambda_2 X_2 = 0, \tag{Ap.2}$$

where $X_1''''$ stands for $\frac{d^4 X_1}{dx^4}$. $\left((\text{Ap.1})\times X_2 - (\text{Ap.2})\times X_1\right)$ results in

$$(\lambda_2 - \lambda_1)X_1 X_2 + \frac{EJ}{\rho S}(X_1'''' X_2 - X_2'''' X_1) + \frac{T}{\rho S}(X_2'' X_1 - X_1'' X_2) = 0. \tag{Ap.3}$$

Thus

$$\int_0^L X_1 X_2 \, dx = \frac{T}{\rho S(\lambda_2 - \lambda_1)} \int_0^L (X_1'' X_2 - X_2'' X_1) \, dx - \frac{EJ}{\rho S(\lambda_2 - \lambda_1)} \int_0^L (X_1'''' X_2 - X_2'''' X_1) \, dx. \tag{Ap.4}$$

Since

$$(X_1'' X_2 - X_2'' X_1) \, dx = d(X_1' X_2 - X_2' X_1), \tag{Ap.5}$$

the first term in Eq. (Ap.4) equals to zero from boundary conditions. Similarly,

$$(X_1'''' X_2 - X_2'''' X_1) \, dx = d(X_1''' X_2 - X_2''' X_1 + X_2'' X_1' - X_1'' X_2'), \tag{Ap.6}$$

so the second term in Eq. (Ap.4) also vanishes. Therefore, eigenfunctions of different eigenvalues of Eq. (2.3) are orthogonal. Q.E.D.

## References

[1] L. Hiller, P. Ruiz, "Synthesizing musical sounds by solving the wave equation for vibrating objects," J. Audio Eng Soc. **19**, 462 (1971)

[2] A. Chaigne andA. Askenfelt, "Numerical simulations of piano strings. Ⅰ. A physical model for a struck string using finite difference methods," J. Acoust. Soc. Am. **95**, 1112 (1994).

[3] A. Chaigne and A. Askenfelt, "Numerical simulations of struck strings. II. Comparisons with measurements and systematic exploration of some hammer-string parameters," J. Acoust. Soc. Am. **95**, 1631 (1994)

[4] A. Chaigne, "On the use of finite differences for musical synthesis. Application to plucked stringed instruments," J. Acoust. **5**, 181 (1992)

[5] H. A. Conklin, Jr., "Design and tone in the mechanoacoustic piano. Part I. Piano hammers and tonal effects", J. Acoust. Soc. Am. **99**, 3286 (1996)

[6] Xianbin Jin, Gaokun Feng, Guifang Lu. *Structure and Manufacture of Piano* (China Light Industry Press, Beijing, 2002, in Chinese).